# Machine Learning and Seismic Attributes for Petroleum Prospect Generation and Evaluation: An Example from Offshore Australia

Mohammed Farfour, Rachid Hedjam, Douglas Foster, Said Gaci

**Corresponding author:** Mohammed Farfour, mfarfour@squ.edu.om

**Abstract**

The presence of hydrocarbons in subsurface rock can manifest in seismic data as abnormal anomalies of amplitudes frequencies, or vertical noise resembling chimneys that crosscut overlying reflections. Seismic attributes play pivotal role in detecting such anomalies. The growing number of seismic and elastic attributes poses a challenge, making the full benefit from each attribute very difficult, if not impossible. Various approaches are routinely employed to select the best attributes for specific purposes. Machine learning (ML) algorithms have demonstrated good capabilities in combining appropriate attributes to address reservoir characterization problems. This study aims to use and combine seismic and elastic attributes to detect hydrocarbon-saturated reservoirs, source rock, and seal rocks in the Poseidon field, Offshore Australia. A large number of attributes are extracted from seismic data and from impedance data. Artificial Neural Networks (ANN) are implemented to combine the extracted attributes and convert them into Resistivity volume, and Gamma Ray volume from which Shale probability volume, Sand volume probability volume, Effective Porosity volume, and Gas Chimney Probability Cubes. The cubes are deployed for a detailed analysis of the petroleum system in the area. The produced Shale volume and Resistivity cube helped delineate the seal rock and source rock in the area. Next, the reservoir intervals were identified using Porosity, Shale, and Resistivity volumes. A pre-trained Convolutional Neural Network (CNN) is trained using another carefully selected attribute set to detect subtle faults that hydrocarbons might migrated through from source rock to trap. The integration of all the extracted cubes contributed to find new prospects in the area and assess their geological probability of success. The proposed approach stands out for its multi-physical attribute integration, Machine Learning and Human expertise incorporation, possible applicability to other fields.

**Introduction**

Geological formations are distinguished by their unique elastic properties which are essential understanding subsurface structures and formations. Seismic data, ideally, showcase changes or contrasts in the elastic properties of rocks; however, the convolution with the source wavelet introduces limitations in resolution and ambiguity. Seismic attributes are defines as any information derived from seismic data. Seismic attributes play a crucial role in facilitating the extraction of physical and geometrical features from seismic data, hence, enhancing our understanding of complex geological settings and conditions. The significance of seismic attributes extends to their ability to uncover hidden structural features, and detect fluid expressions of subsurface formations (Farfour et al. 2012a; Farfour et al. 2012b; Farfour et al. 2014, Farfour et al. 2016; Farfour et al. 2017). Seismic inversion is a process that enables interpreters to derive acoustic and elastic impedance data from seismic traces. Elastic attributes are known any quantity computed from impedances. In fact, from impedances, one can extract a variety of elastic attributes that can be closely linked to fluid and lithology of geological formations.

With the consistently increasing number of seismic attributes, selecting the appropriate attributes for specific objectives poses a challenge in modern seismic interpretation. Ideally, seismic and elastic attributes serve as dimensions in which seismic objects can be presented and projected. Various attempts and studies in the literature highlight the benefits of multi-attribute sets, frequently combined through diverse methodologies. Machine learning (ML), including both shallow and deep learning approaches, are increasingly applied to determine the minimum order of dimensionality necessary for the effective detection and characterization of reservoirs (Farfour et al. 2012; Dixit and Mandal, 2020; Ismail et al. 2022). This reduction in dimensionality not only saves significant computation time and enhances computational efficiency but also encourages interpreters to adopt ML-based methods across various interpretation tasks and operations, such as pattern recognition, facies classification, and lithology predictions.

In the realm of reservoir characterization and petroleum system analysis, ML applications aims to improve the accuracy and efficiency of analyzing and characterizing critical petroleum system elements such as reservoir rocks, source rocks, and seal rocks. Challenges in formation characterization stem from two main sources: 1) the extraction of geological information from geophysical data which is not unique in nature 2) the diversity of the data which are distinct in scale, range, and nature. Traditional methods of reservoir characterization rely heavily on the expert's interpretation of seismic data, well logs, and geological insights of the area, requiring significant manual integration of the resulting products. Recent advancements in ML-based approaches introduced tools that enable interpreters to extract essential features and information directly from large and complex datasets, thereby reducing reliance on expert work. Moreover, these advances helped facilitate the integration of extracted information in more practical manner compared to traditional methods.

The growing number of successful applications of ML algorithms in reservoir geophysics demonstrates their vital role in characterizing and understanding reservoirs. These applications are from many different world basins. For instance, Farfour et al. (2012a) have utilized Artificial Neural Network to detect shallow gas-saturated zones from F3 block, North Sea. They combined various seismic attributes including instantaneous attributes, spectral decomposition components at different frequencies to detect anomalies associated with shallow gas sands. Zhao, et al. (2015) have used several seismic facies classification algorithms including k-means, self-organizing maps, generative topographic mapping, support vector machines, Gaussian mixture models, and artificial neural networks to extract several geological features from seismic data from Canterbury Basin, offshore New Zeeland. Wrona et al. (2017) have trained a range of ML models to conduct seismic interpretation and automatic facies prediction based on a comprehensive list of seismic attributes. Dixit and Mandal (2020) combined a list of seismic attributes extracted from data from offshore Australia to generate gas chimney probability cube to help track fluid migration and leakage of deep-seated gas-reservoir. Ismail et al. (2022) have deployed ANN on seismic data from offshore Nile Delta to combine seismic attributes and use them to delineate a gas channel reservoir and gas chimneys associated with the gas migration.

In this study, different ML approaches, including Shallow Neural Networks and Deep Neural networks, are utilized to identify prospects in the Poseidon field, Offshore Australia. Shallow Learning approaches are applied to detect fluid expressions and gas migration pathways, as well as to create Gamma-Ray cube and Resistivity cube, which assist in identifying different petroleum system elements (source rock, reservoir rock, and seal rock). The deep learning Convolutional Neural Network (CNN) is deployed to generate detailed fault trends in the zone of interest. Results from all generated cubes are integrated to identify new potential hydrocarbon zones and assess their geological risks.

**Artificial Neural Network (ANN): Multilayer Perceptron (MLP)**

ANN are a sort of ML models aiming to learn the relationship between data input and data outputs. The input are represented as vectors of attributes extracted from observed data such as frequencies, amplitudes, etc. of a signal. The outputs describe in general the categories or the classes or the types of the data inputs. They are called also labels. For instance, in the problem of anomaly detection the output can be normal or anomaly whether the input data is a normal signal (known) or abnormal one (not known). The model is trained based on a set of training samples (vectors of attributes) and their respective labels in order to adjust the weight of the models to minimize the difference between the predicted labels and the real labels in terms of a predefined loss criterion. Once the model is trained, it will be exposed to unseen samples (those not used in the training step) to evaluate its performance. The first phase is called the training phase and the second one is called test phase. Figure 1 shows a typical MLP ANN with multiple layers, in which each layers is composed by a set of nodes. Each node in a given layer acts as a nonlinear function transforming the input data of all nodes in the previous layer. Therefore, the input data is passed to the ANN and moves through the layers and undergoes multiple nonlinear transformations until the final output or corresponding predicted label is produced. The predicted labels are then compared to the true labels to calculate the total error $E$ in terms of root mean square error, which in turn back-propagates to the first layers to update all network weights.

$$E = \frac{1}{2}\sum_{i=1}^{n}(y_i - O_i)^2 \quad (1)$$

where, $y_i$ and $O_i$ are respectively the true and predicted labels of the $i^{th}$ training sample; and $n$ is total number of the training samples. At iteration $(t + 1)$ The back-propagation algorithm updates the weights of the networks, $w_{ij}$, as follows:

$$w_{ij}^{(t+1)} = w_{ij}^{(t)} + \Delta w_i^{(t)} \quad (2)$$

where, $\Delta w_{ij}^{(t)}$ is the change in weight, which in turn can be calculated as follows:

$$\Delta w_{ij}^{(t)} = -\eta \frac{\partial E}{\partial w_{ij}} \quad (3)$$

where, $\eta$ is the learning rate predefined by the user.

In a multi-class classification problem, for each input the ANNs produce a vector $O = (o_1, .. o_k, ... , o_K)$, where $K$ is the total number of available classes, and $o_k$ is the output value of class $k^{th}$. Since the $o_k$ value can be any float number, the ANN transforms (scales) it to a value in a specific range using a nonlinear activation function. There are many types of activation functions and the best known (in particular for multi-layer perceptron neural nets) is the Sigmoid function that transforms any float value to a values in the range [0..1]. This showed a very significant impact on the stability and robustness of ANNs. The Sigmoid function applied on $o_i$ has the following form: $z_i = \frac{1}{1+e^{-o_i}}$

Finally, the ANN uses in general the softmax function to activate the output vector $(z_1, .. z_k, .. z_K)$ and decide about the label to predict for a given input sample. The softmax function computes the relative probabilities of each possible class label and the one with the highest probability will be predicted. Hence, the softmax function $S$ of the output vector is computed as follows:

$$S(z_k) = \frac{e^{z_k}}{\sum_{l=1}^{K} e^{z_l}} \quad (3)$$

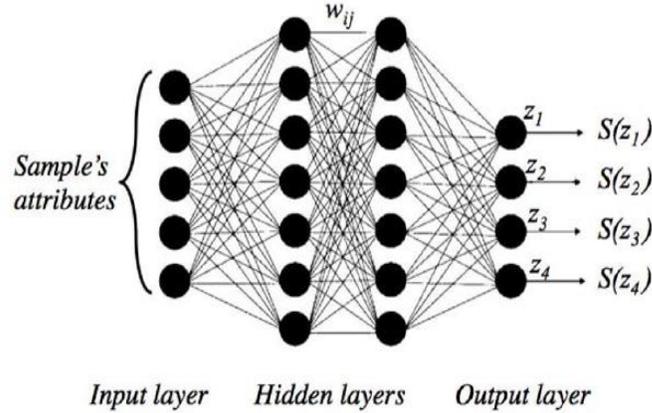

**Figure 1.** A typical architecture of an ANN with 2 hidden layers, one input and one output layer.

Various artificial neural network (ANN) algorithms are employed in hydrocarbon exploration, which are chosen based on the nature of the problem and available input datasets (Farfour et al. 2012).

**Convolutional Neural Network (CNN)**

Convolutional Neural Networks (CNNs) represent a specialized class of Deep Learning architectures tailored for processing structured grid data, notably images. In contrast to Multilayer Perceptron (MLP), which treat input data as flat vectors, CNNs maintain the spatial organization of the input through convolutional layers. These layers employ learnable filters to detect patterns across localized regions of the input, fostering the automatic extraction of hierarchical features. Additionally, CNNs incorporate pooling layers to downsample feature maps, reducing computational complexity while retaining essential information. Typically, CNN architectures comprise alternating convolutional and pooling layers, culminating in fully connected layers for classification tasks. This design enables CNNs to effectively capture spatial hierarchies and translational invariance in data, rendering them particularly adept at tasks like image classification, object detection, and image segmentation. Conversely, MLPs struggle with structured grid data due to their inability to preserve spatial relationships, resulting in diminished performance for tasks involving image recognition.

CNNs have also found extensive applications in processing one-dimensional (1D) data, leading to the development of specialized architectures known as 1D CNNs (see Figure 2). These models have been effectively employed in tasks involving sequential data such as time series analysis, audio signal processing, and natural language processing. In essence, the term "1D CNN" reflects the adaptation of convolutional neural networks to capture temporal patterns and dependencies inherent in one-dimensional sequences, expanding their utility beyond traditional image processing tasks.

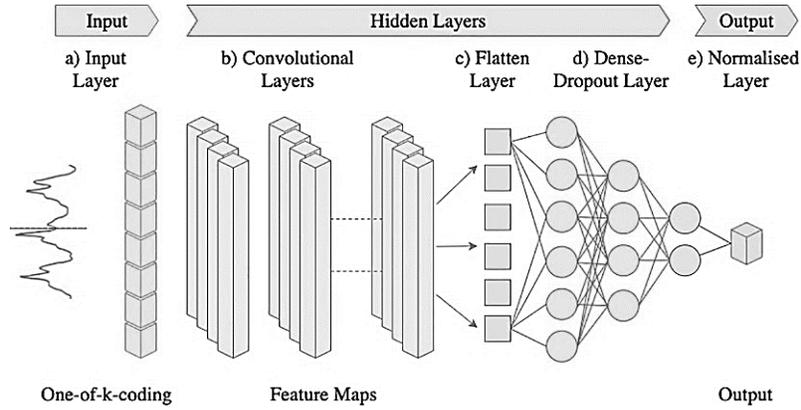

**Figure 2.** Convolutional neural networks (CNN) and one-dimension convolutional neural network (1D-CNN) architecture (Sánchez-Reolid et al. 2022)

In a convolutional layer, the output of a convolutional layer $O$ is obtained by convolving the input sequence $X$ with a set of learnable filters $W$, applying a bias term $b$, and an activation function $f$:

$$O_i = f(\sum_{j=1}^{K}(X_{i+j-1} * X_j) + b), \quad (4)$$

where $O_i$ is the $ith$ element of the output feature map, $X_{i+j-1}$ denotes the $jth$ element of the input sequence centered around position $i$, $W_j$ represents the $jth$ filter, $K$ is the filter size, $*$ denotes the convolution operation, and $b$ is the bias term.

In a pooling layer, the output feature map $O$ is obtained by down sampling the input feature map $I$ using a pooling function $g$:

$$O_i = g(I), \quad (5)$$

Common pooling functions include max pooling and average pooling, where the maximum or average value within each pooling region is retained, respectively.

In a fully connected layer, the processing is performed the same way as for MLP discussed in the previous section.

**Methodology:**

In this study, seismic amplitude-related attributes, including instantaneous attributes, Scaled Poisson reflectivity, AVO gradient, and their products are utilized to detect hydrocarbon-saturated reservoirs in the Poseidon field, North Western Australia (Farfour and Foster, 2021). Geometrical attributes, such as similarity and coherence, are then employed to identify faults and vertical chaotic noises associated with hydrocarbon migration. Next, seismic inversion is used to convert prestack gathers to elastic attributes such as P-impedance, S-impedance, and density from which various elastic attributes are derived. The extracted attributes include Vp-Vs ratio, Poisson ratio, Bulk modulus, Shear modulus, Lambda-Rho, Mu-Rho. Table 1 lists all extracted elastic parameters and how they are extracted from P-wave and S-wave impedances. To avoid complications and uncertainties related to the separation of the density from velocities, it is a common practice to use the expressions in terms of impedances.

Table 1: Elastic parameters derived from Vp, Vs, and density.

| Parameter | Real expression | Attributes from impedances |
|---|---|---|
| **Bulk modulus (incompressibility)** | $\kappa = \rho V_p^2 - \frac{4}{3}\rho V_s^2$ | $\rho\kappa = (\rho V_p)^2 - \frac{4}{3}(\rho V_s)^2$ |
| **1st Lame parameter** | $\lambda = \rho V_p^2 - 2\rho V_s^2$ | $\rho\lambda = (\rho V_p)^2 - 2(\rho V_s)^2$ |
| **2nd Lame parameter (rigidity)** | $\mu = \rho V_s^2$ | $\rho\mu = (\rho V_s)^2$ |
| **Bulk modulus shear modulus difference** | $\kappa - \mu = \rho V_p^2 - \frac{1}{3}\rho V_s^2$ | $\rho(\kappa - \mu) = (\rho V_p)^2 - \frac{1}{3}(\rho V_s)^2$ |
| **Poisson ratio** | $\sigma = \frac{V_p^2 - 2V_s^2}{2V_p^2 - 2V_s^2}$ | $\sigma = \frac{V_p^2 - 2V_s^2}{2V_p^2 - 2V_s^2}$ |
| **Young modulus (stiffness modulus)** | $E = \rho V_s^2 \frac{3V_p^2 - 4V_s^2}{V_p^2 - V_s^2}$ | $\rho E = \frac{3V_p^2 - 4V_s^2}{V_p^2 - V_s^2}(\rho V_s)^2$ |

The above elastic attributes are combined with additional features using a multi-layer Feedforward Neural Network to predict a Gamma Ray model from Gamma Ray logs and Resistivity model from Resistivity logs. From the extracted cubes, facies of the source rock, reservoir rock, and seal rock facies are generated. The chart in Figure 3 illustrates the workflow we used.

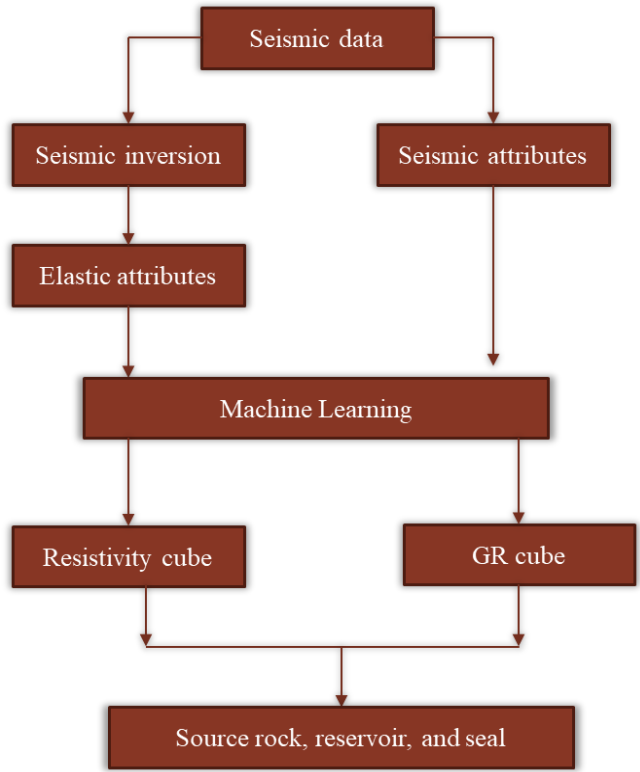

**Figure 3.** A chart showing the workflow used to produce reservoir, source, and seal rock facies from seismic data using ML algorithms.

Additionally, a pre-trained Convolutional Neural Network (CNN), U-Net, is used to predict faults using a distinct combination of seismic attributes. The CNN is heavily pre-trained on synthetic seismic data, eliminating the need for retraining.

The results from the above tasks are integrated to identify prospects and assess associated risks. Probability of geological success (Pg), as defined by Rose (2004), represents the likelihood of finding producible hydrocarbons in a reservoir. This percentage value is commonly used for risked prospective resource calculations (White, 1993) and forms the initial step in a series of investment decisions.

It is commonly accepted that for a geological structure to contain movable hydrocarbons, there must be:

1. Structure (closure, geometry, including stratigraphic) that accumulates and contains the hydrocarbons;
2. Porous reservoir which is thick and permeable enough to allow the fluid to flow;
3. Hydrocarbon fluids that were expelled from mature source rocks and then moved into the reservoir along migration pathways;

4. Seal that keeps the petroleum within the trap.

Explorers extract various geological factors to help them assess the risk of drillable potential zones. Each of these factors is considered equally important, and failure in any one of them implies the absence of hydrocarbons trapped within the evaluated prospect (Nosjean et al., 2021). The risk assessment process of proepects involves estimating these factors without considering seismic amplitude anomalies as direct hydrocarbon indicators. Bacon and Simm (2014) proposed an approach to combine the risk values with direct hydrocarbon indicator (DHI) occurrence from the seismic amplitude using Bayes' theorem through the following equation:

$$P(hc \mid dhi) = \frac{P(dhi|hc)P(hc)}{P(\text{dhi}|hc)P(hc)+P(\text{dhi } nohc)P(\text{nohc})} \quad (6)$$

Where P(hc|dhi) is the probability of hydrocarbons given the observed DHI, P(dhi|hc) the probability of occurrence of the observed DHI if hydrocarbons are present, P( dhi|nohc) the probability of occurrence of the observed DHI if no hydrocarbons are present, P(hc) the a priori probability of hydrocarbons ignoring the DHI (i.e. the 'geological' chance of success), P(dhi│nohc)=1-P(hc) is the a priori probability of no hydrocarbons.

**Results and Discussion**

Elastic and seismic attributes computed at the well locations demonstrated that hydrocarbon-saturated reservoirs can be detected (Figure 4). Vp/Vs, Poisson ratio, K-Mu, and AVO fluid factor are displayed in Figure 4. Clear drops are noticed in Vp/Vs, Poisson ratio, and K-Rho-Mu-Rho at the reservoir interval, which are attributed to the gas presence (Avseth et al. 2015). The AVO fluid factor demonstrated a negative amplitude anomaly which is also a good indication of gas presence (Foster et al. 2010). Before applying attributes, computation or combination using any ML approaches, the data were enhanced by applying structure-oriented filters, specifically the dip-steered median filter. Following data preconditioning, several attributes were utilized to scan the data for potential fluid expressions. The attributes used include: spectral decomposition, instantaneous attributes, energy attribute, SPR-Gradient product (Farfour and Foster, 2022), and Amplitude Component Analysis (Farfour, 2020). In Figure 5, the SPR-Gradient product is presented, displaying clear anomalies associated with hydrocarbons observed at well location, precisely at the reservoir level (green). Subsequently, seismic prestack inversion was conducted to extract different impedance-based attributes. The list of extracted attributes involves P-impedance, S-impedance, Vp toVs ratio, Lamda-Rho, Poisson ratio. In Figure 6, a section of K-Rho-Mu-Rho along the same line of SPR-Gradient. Gas-saturated sands demonstrate lower values of bulk modulus (K-Mu) as a result of their compressibility.

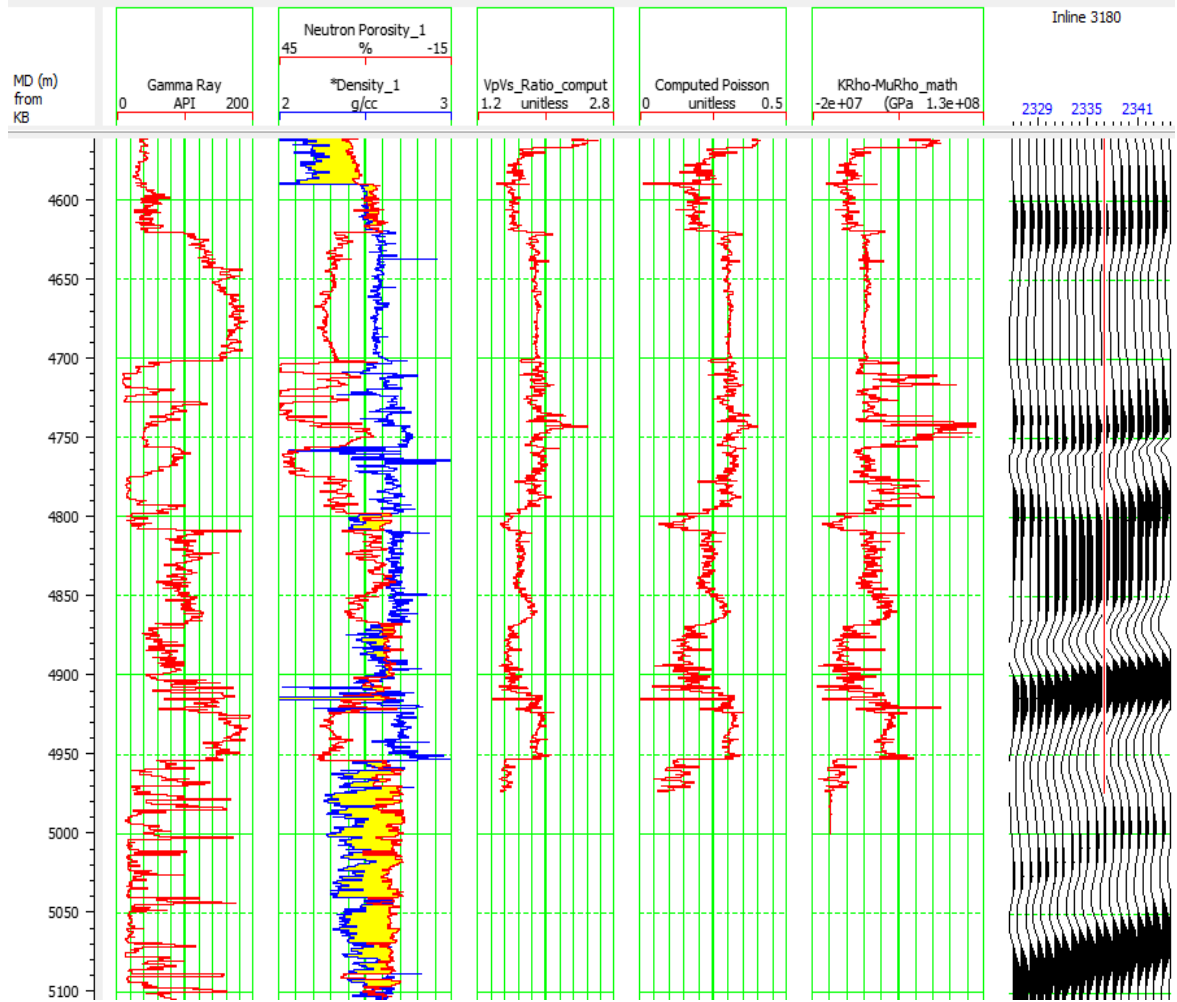

**Figure 4.** Well logs of GR, Neutron Porosity and density, Vp/Vs ratio, Poisson ratio, and a seismic fluid factor from AVO intercept and gradient.

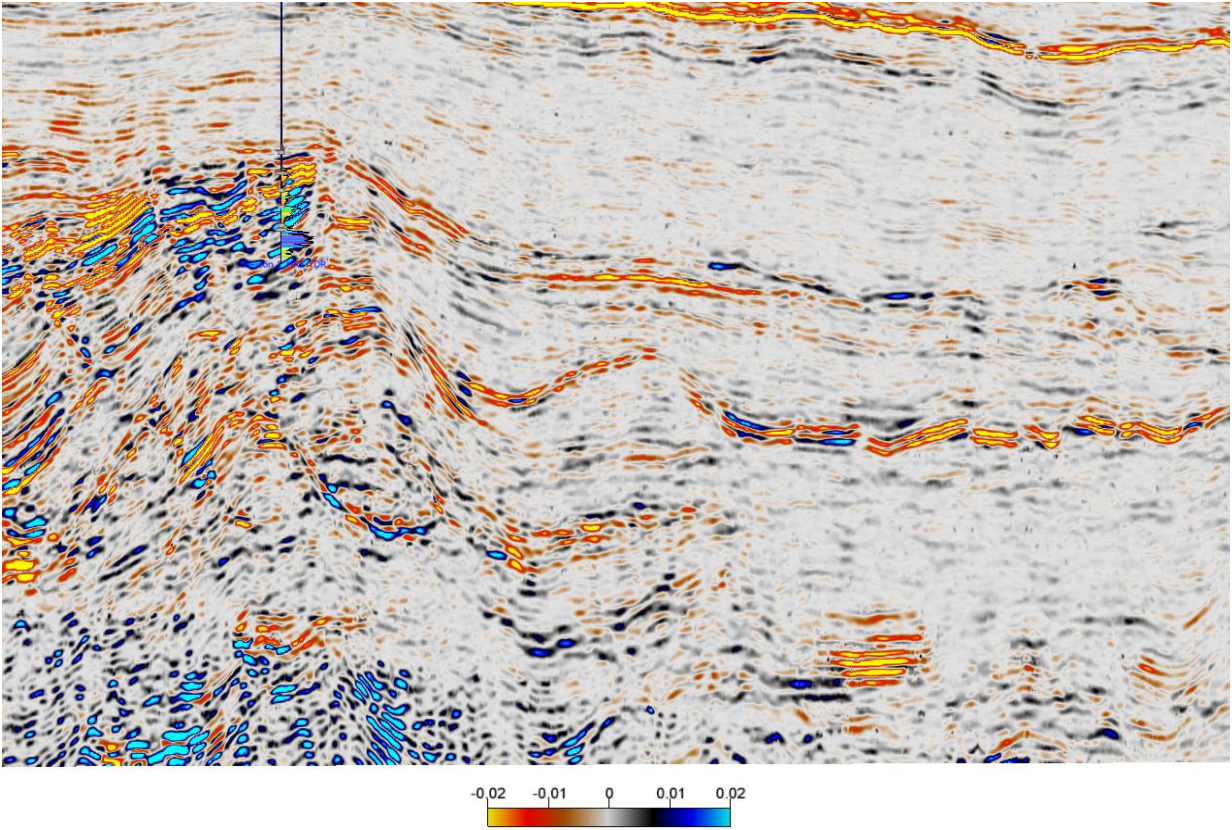

**Figure 5.** SPRxGradient section extracted at a line passing by the discovery well, Poseidon 1. Reservoir top and bottom distinguish themselves from the background and appear in cyan.

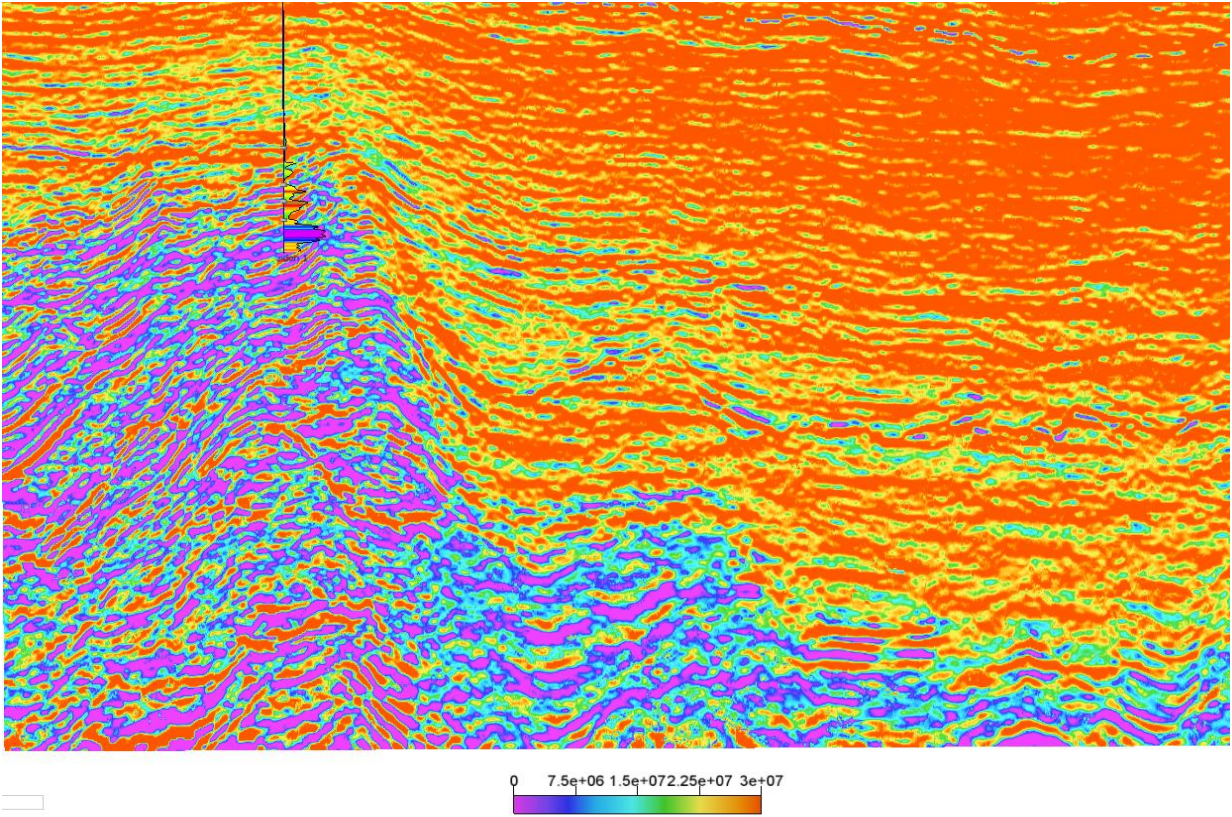

**Figure 6.** KRho-MuRho section computed the inverted P-impedance, S-impedance passing through the discovery well Poseidon 1. The reservoir zone show low amplitude anomalies.

Figure 7 displays a screenshot of the neural network training process to predict the Gamma Ray and the Resistivity models from the combination of seismic attributes and elastic attributes. The RMS error initially starts high and then decreases to a stable level, indicating the successful establishment of a relationship between the attribute set and the target property. The final products from the training are the probability cubes that represent the natural radioactivity (GR) and the electrical resistivity of the geological formations.

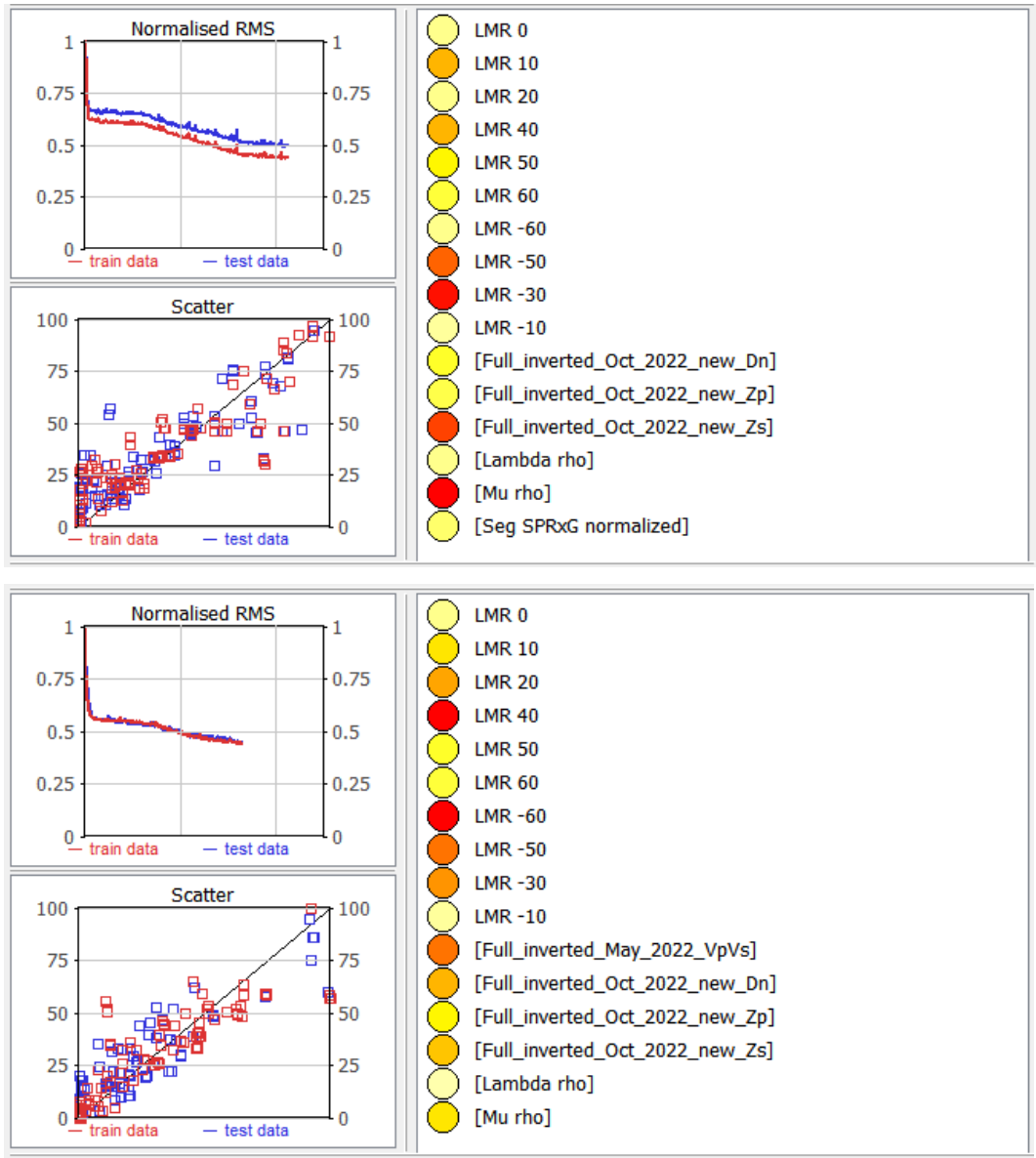

**Figure 7.** Training process for GR prediction (top) and Resistivity prediction (bottom). Lambda-Rho is computed at different time samples and used along with P-impedance, S-impedance, Vp/Vs, and SPR-Gradient.

The predicted Gamma Ray and Resistivity models, shown in Figure 8, indicate the presence of source rocks, reservoir rocks, and seal rocks (initially identified at well locations). Due to their organic matter content, source rocks typically exhibit high Gamma Ray values and relatively higher resistivity compared to seal rocks. A resistivity cutoff of 10 ohm·m was applied to discriminate between seal and source rocks. Ultra-thin source rock intervals were identified at 4702 m and 4924 m using Rock-Eval Pyrolysis and (ConocoPhillips, 2011). The latter intervals were used for calibration. Reservoirs, on the other hand, are characterized by their high sand volume, effective porosity, and elevated resistivity. We use 60 API as the cutoff between shale and sand formations, a resistivity of 100 ohm·m, and 15% porosity as a cutoff to distinguish reservoirs from non-reservoir formations. These values were calibrated using data from the Poseidon 1 well completion report. A good agreement is observed between the extracted volumes and the well logs and tops.

The predicted reservoir rocks demonstrate a good continuity at some intervals that is disrupted by the tectonic deformations resulting from the extensional regime in the area. Some discontinuities are found to be due to depositional changes.

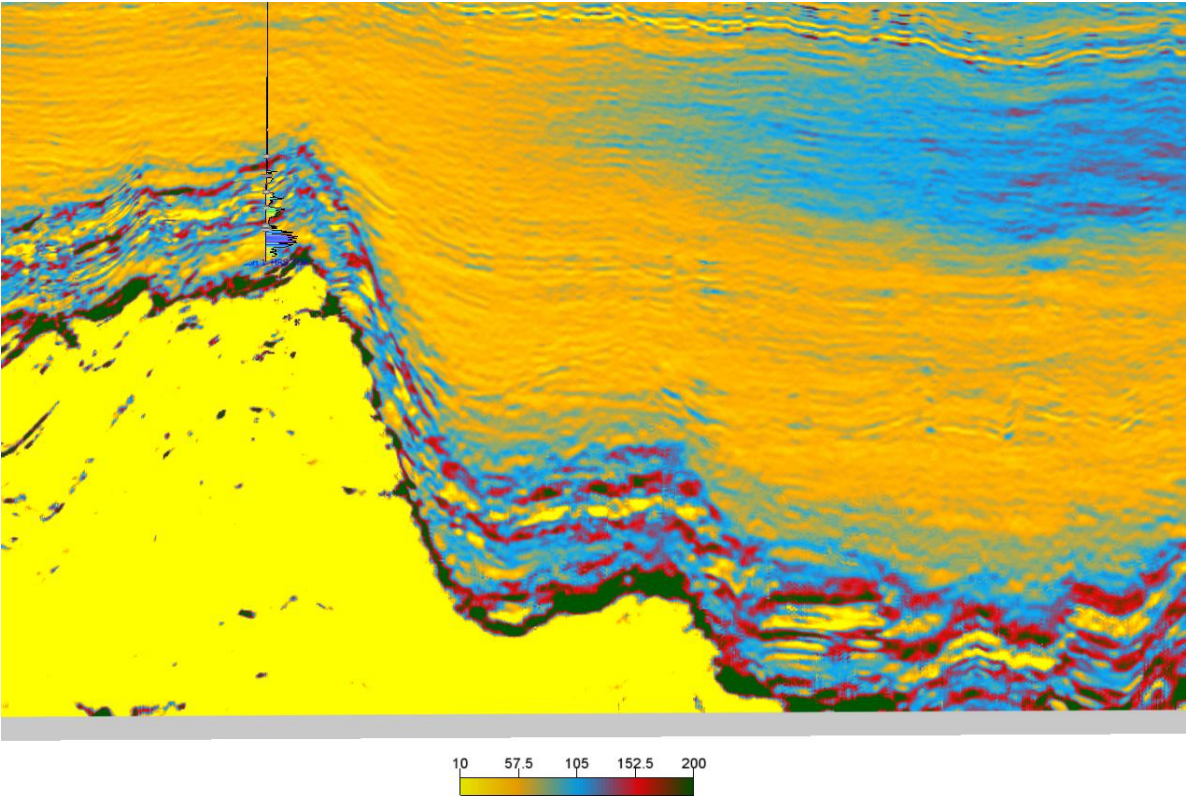

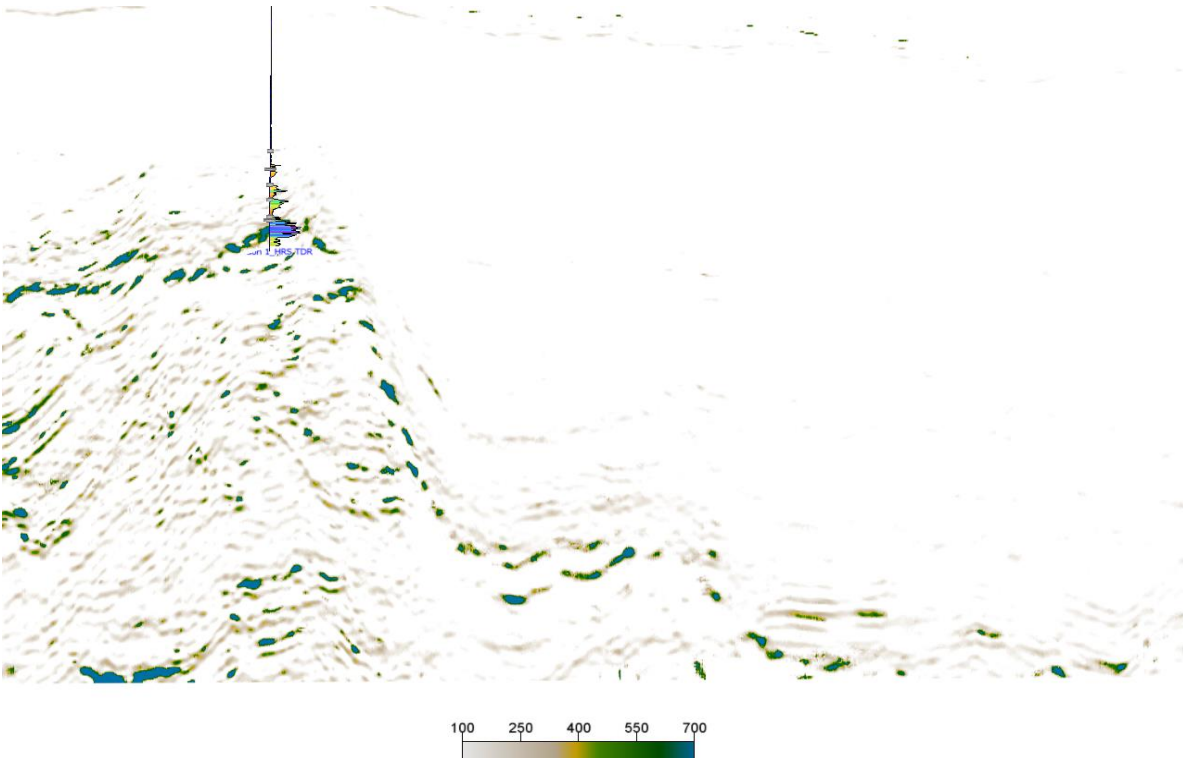

**Figure 8.** Predicted Gamma Ray (top) and Resistivity data (bottom) showing reservoir zones with high resistivity and low GR. Seal, and source rocks are showing high GR but low resistivity.

In Figure 9, we integrate all the facies together, source, reservoir, and seal facies and show them on the same cube.

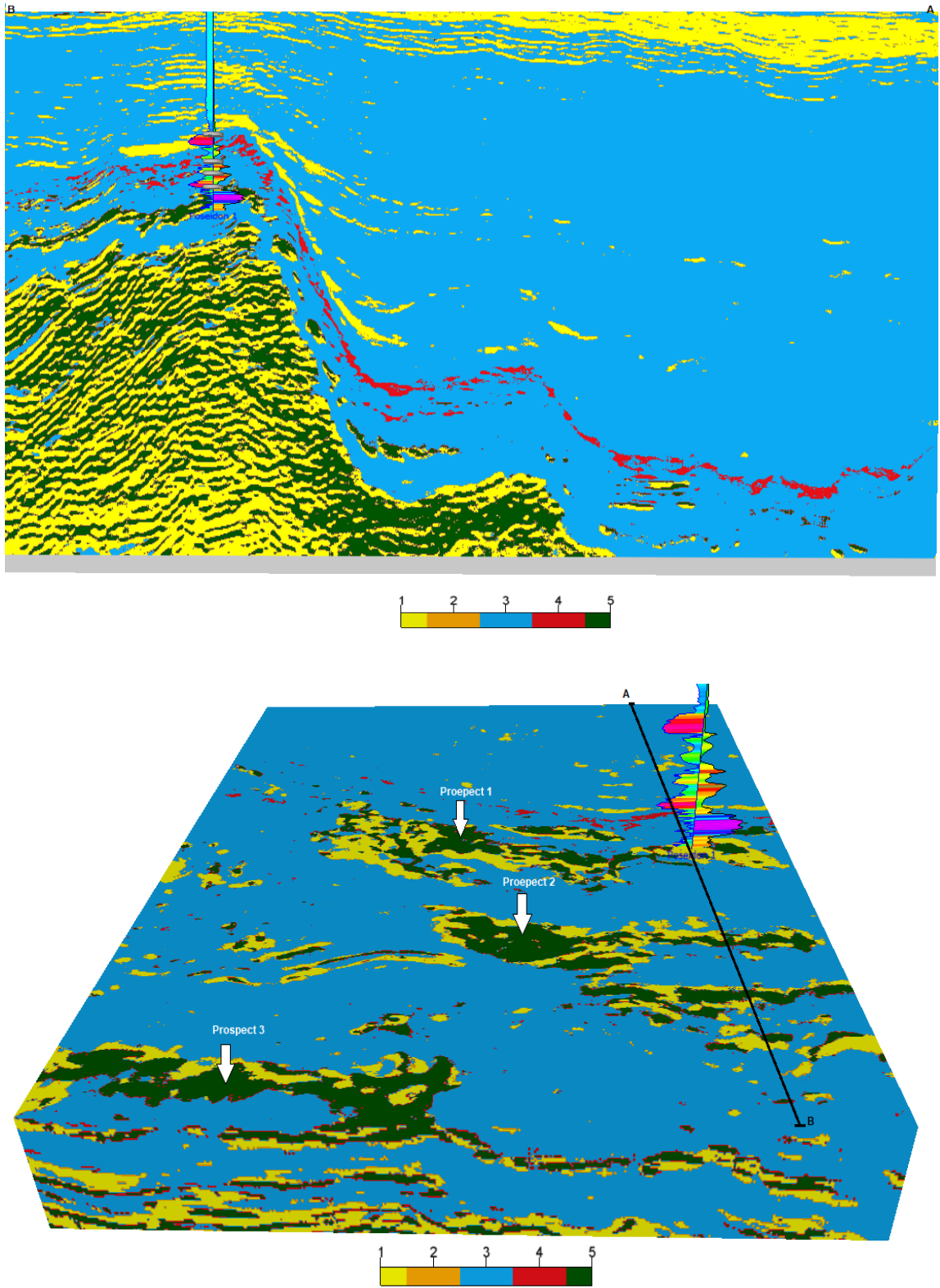

**Figure 9.** The three facies are combined in a 3-D model. The model helped identify new prospective zones. Reservoir is in green, source in read, seal rock is in blue.

Furthermore, another set of attributes was used afterwards to train the network to predict possible fluid migration pathways. Hydrocarbon migration pathways appear in seismic data as chaotic noise that locally disturbs the continuity of seismic events. At this stage, a binary classification model was built to categorize the data samples in the 3-D model into two (02) classes related to the presence/ absence of chimney. The identified noise was introduced to the neural network, and after several attempts, a stable minimum RMS error was achieved, defining the relationship between chimneys and the attribute set. It is worth noting that several authors documented the success of ANN in detecting gas chimneys and fluid migration paths (Singh et al., 2016; Dixit and Mandal, 2020; Ismail et al. 2022). The above examples are from many different world basins.

Lastly, the U-Net was invoked to help us predict subtle fault trends that may not be easy to track manually. The U-Net is already trained using some geometrical attributes to detect such faults. A very detailed fault cube was generated and calibrated with faults from manual interpretation. Combining the fault probability cube with the chimney cube revealed probable migration paths and demonstrated the capacity of the seal to stop migrating fluid (Figure 10).

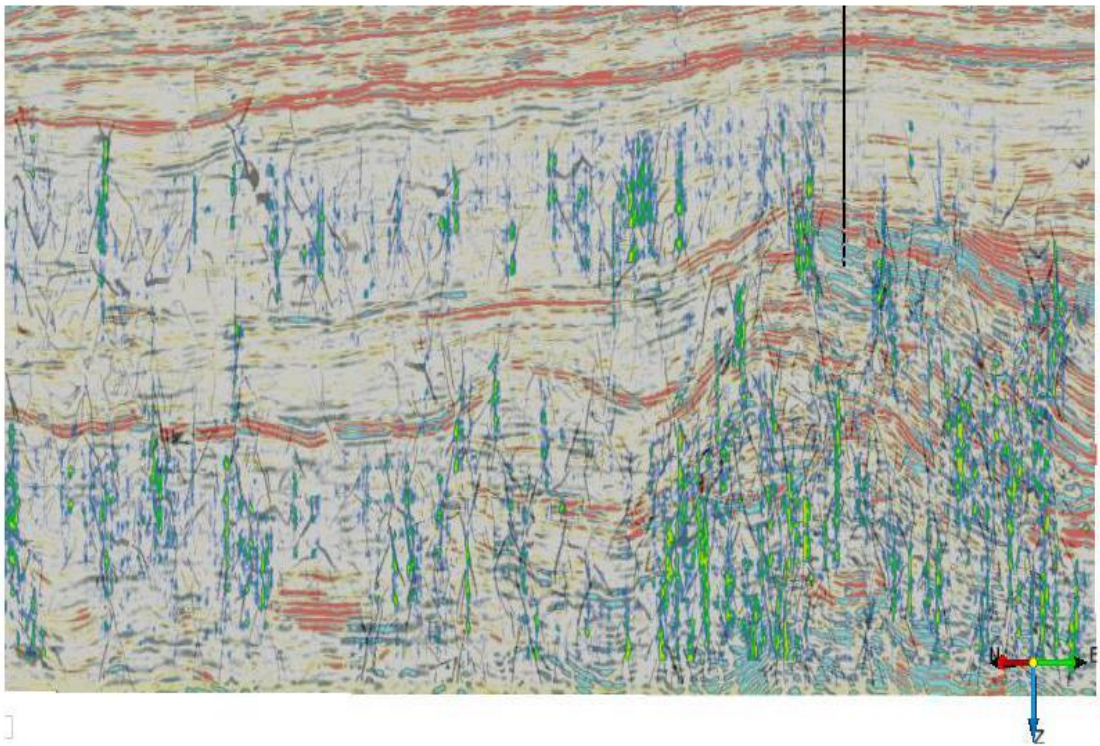

**Figure 10.** Gas chimney model co-blended with fault model predicted by CNN.

Using the maps of the predicted facies (source, reservoir, and seal) and chimney cube from the ANN, the faults from the CNN, and the geological prior-knowledge gained from petroleum system elements in the area, new prospects were identified, and probability of success values were assigned to each prospect as detailed in Table 02. The analysis of risk factors Kg-1, Kg-2, Kg-3, Kg-4, Kg-5 related to Reservoir, Trap, Seal, Source, Timing respectively were analyzed, indicating that most of the risk comes from the seal. Initial probability of success PoS0 is computed directly from the product of the risk factors. The computed probabilities were then integrated with the observations from ANN using the equation 1. The new probability rates of success (PoS1) is presented in the same table (Table 02).

Table 02: The chance of success values assigned for the selected prospects 1, 2, and 3 in Figure 9.

|  | Prospect 1 | Prospect 2 | Propsect 3 |
| --- | --- | --- | --- |
| Kg-1 | 0.90 | 0.90 | 0.90 |
| Kg-2 | 0.90 | 0.90 | 0.90 |
| Kg-3 | 0.80 | 0.80 | 0.90 |
| Kg-4 | 0.90 | 0.80 | 0.70 |
| Kg-5 | 0.80 | 0.80 | 0.80 |
| PoS0 | 0.47 | 0.41 | 0.41 |
| PoS1 | 0.78 | 0.74 | 0.73 |

It is evident from the table that including the probability of fluid expression from the ANN has significantly enhanced the chances of success.

It is worth mentioning that the risk assessment is always challenging task and prone to a great deal of subjectivity, as there is no exact equation to calculate the geological risks consistently, precisely, and accurately (Milkov, 2021). Factors influencing these assessments can vary significantly from a person to another.

**Conclusion:**

In this study, ML approaches were employed to combine seismic and elastic attributes to detect hydrocarbon zones, produce resistivity and Gamma Ray cubes, and predict gas chimneys. The computed cubes helped delineate and analyze the petroleum system in area, source rock, seal rock, and reservoir rock. Additionally, a deep neural network, convolutional neural network, was utilized to predict subtle faults responsible for the hydrocarbon migration and caused by tectonic deformations throughout the history of the basin. The outcomes from these ML-based methods were collectively utilized to identify new prospective zones and assess their associated risks.

**References**

- Avseth, Per and Tor Veggeland. 2015. Seismic Screening of Rock Stiffness and Fluid Softening Using Rock-Physics Attributes. Interpretation 3(4):SAE85–93.
- Azadpour, M., Saberi, M. R., Javaherian, A., & Shabani, M. (2020). Rock physics model-based prediction of shear wave velocity utilizing machine learning technique for a carbonate reservoir, southwest Iran. Journal of Petroleum Science and Engineering, 195, 107864. https://doi.org/10.1016/j.petrol.2020.107864
- Bacon, M, and Simm, R., 2014. Seismic Amplitude: An Interpreter's Handbook. Cambridge University Press, 283p. https://doi.org/10.1017/CBO9780511984501.
- ConocoPhillips (2011). Poseidon-1 well completion report, volume 2: Interpretive data. Document No: POSE-1/004.
- Dixit, A., & Mandal, A. (2020). Detection of gas chimney and its linkage with deep-seated reservoir in Poseidon, NW shelf, Australia from 3D seismic data using multi-attribute analysis and artificial neural network approach. Journal of Natural Gas Science and Engineering, 83. https://doi.org/10.1016/j.jngse.2020.103586
- Farfour, M. (2020). Amplitude components analysis: Theory and application. The Leading Edge, 39(1), 62a1-62a6. https://doi.org/10.1190/tle39010062a1.1
- Farfour, M. and Foster, D. (2021). New AVO expression and attribute based on scaled Poisson reflectivity. Journal of Applied Geophysics, 185, https://doi.org/10.1016/j.jappgeo.2021.104255
- Farfour, M., & Foster, D. (2022). Detection of hydrocarbon- saturated reservoirs in a challenging geological setting using AVO attributes: A case study from Poseidon field, Offshore Northwest region of Australia. Journal of Applied Geophysics, 203, 104687. https://doi.org/10.1016/j.jappgeo.2022.104687.
- Farfour, M., Ferahtia, J., Djarfour, N., & Aitouch, M. A. (2017). Seismic spectral decomposition applications in seismic. Special Publications, 93–113. https://doi.org/10.1002/9781119227519.ch6
- Farfour, M., Yoon, W. J., & Jang, S. (2016). Energy-weighted Amplitude Variation with Offset: A new AVO attribute for low impedance gas sands. Journal of Applied Geophysics, 129, 167–177. https://doi.org/10.1016/j.jappgeo.2016.03.032
- Farfour, M., Yoon, W. J., Ferahtia, J., & Djarfour, N. (2012a). Seismic attributes combination to enhance detection of bright spot associated with hydrocarbons. Geosystem Engineering, 15(3), 143–150. https://doi.org/10.1080/12269328.2012.702089
- Farfour, M., Yoon, W.J. (2014). Ultra-Thin Bed Reservoir Interpretation Using Seismic Attributes. Arab Journal of Science and Engineering 39, 379–386. https://doi.org/10.1007/s13369-013-0866-9

- Farfour, M.,Yoon, W.J. Jo. Y. (2012b). Spectral decomposition in illuminating thin sand channel reservoir, Alberta, Canada. Canadian Journal of Pure Applied Sciences 6, 1981-1990.
- Foster, D. J., R. G. Keys, and F. D. Lane. 2010. Interpretation of AVO Anomalies. Geophysics 75, 5, 1SO-Z116.
- Ismail, A., Ewida, H. F., Nazeri, S., Al-Ibiary, M. G., & Zollo, A. (2022). Gas channels and chimneys prediction using artificial neural networks and multi-seismic attributes, offshore West Nile Delta, Egypt. Journal of Petroleum Science and Engineering, 208. https://doi.org/10.1016/j.petrol.2021.109349
- Milkov, A. V. (2021). Reporting the expected exploration outcome: When, why and how the probability of geological success and success-case volumes for the well differ from those for the prospect. Journal of Petroleum Science and Engineering, 204, 108754. https://doi.org/10.1016/j.petrol.2021.108754
- Nosjean, N., Holeywell, R., Pettingill, H., Roden, R., & Forrest, M. (2021). Geological probability of success assessment for amplitude-driven Prospects: A Nile Delta case study. Journal of Petroleum Science and Engineering, 202, 108515. https://doi.org/10.1016/j.petrol.2021.108515
- Rose, P.R., 2004. Risk analysis and management of petroleum exploration ventures. AAPG methods in Exploration series 12, 164p.
- Sánchez-Reolid R, de la Rosa FL, Lopez MT, Fernández-Caballero A., (2022). One-dimensional convolutional neural networks for low/high arousal classification from electrodermal activity. Biomedical Signal Processing and Control. 71:103203.
- Singh, D, Chinmoy, P., Kalachand, K., 2016. Interpretation of gas chimney from seismic data using artificial neural network: A study from Maari 3D prospect in the Taranaki basin, New Zealand. Journal of Natural Gas Science and Engineering, 36, 339-357.
- Sweere, S.F., Valtchanov, I., Lieu, M., Vojtekova, A., Verdugo, E., Santos-Lleo, M., Pacaud, F., Briassouli, A., Cámpora Pérez, D., 2022. Deep learning-based super-resolution and de-noising for xmm-newton images.
- Wrona, T., Pan, I., Gawthorpe, R. L., & Fossen, H. (2018). Seismic facies analysis using machine learning. Geophysics, 83(5), O83–O95. https://doi.org/10.1190/geo2017-0595.1
- Zhao, T., Jayaram, V., Roy, A., & Marfurt, K. J. (2015). A comparison of classification techniques for seismic facies recognition. Interpretation, 3(4), SAE29–SAE58. https://doi.org/10.1190/INT-2015-0044.1.20